\documentclass[12pt,epsfig]{article}
\usepackage{graphicx,amsmath,amssymb}

\newlength{\dinwidth}
\newlength{\dinmargin}
\def\lapproxeq{\lower .7ex\hbox{$\;\stackrel{\textstyle
<}{\sim}\;$}}
\def\gapproxeq{\lower .7ex\hbox{$\;\stackrel{\textstyle
>}{\sim}\;$}}
\def\gtrsim{\lower .7ex\hbox{$\;\stackrel{\textstyle
>}{\sim}\;$}}
\def\lesim{\lower .7ex\hbox{$\;\stackrel{\textstyle
<}{\sim}\;$}}

\def\be{\begin{equation}}
\def\ee{\end{equation}}
\def\bea{\begin{eqnarray}}
\def\eea{\end{eqnarray}}

\def\qq{q\bar{q}}

\def\ra{ \rightarrow }

\def\GeV{\rm GeV}

\begin{document}
\begin{flushright}
IPPP/06/10 \\
DCPT/06/20 \\
22nd February 2006 \\

\end{flushright}

\vspace*{0.5cm}

\begin{center}
{\Large \bf Leading neutron spectra}

\vspace*{1cm}
\textsc{A.B.~Kaidalov$^{a,b}$, V.A.~Khoze$^{b,c}$, A.D. Martin$^b$ and M.G. Ryskin$^{b,c}$} \\

\vspace*{0.5cm}
$^a$ Institute of Theoretical and Experimental Physics, Moscow, 117259, Russia\\
$^b$ Department of Physics and Institute for
Particle Physics Phenomenology, \\
University of Durham, DH1 3LE, UK \\
$^c$ Petersburg Nuclear Physics Institute, Gatchina,
St.~Petersburg, 188300, Russia \\

\end{center}

\vspace*{0.5cm}

\begin{abstract}
It is shown that the observation of the spectra of leading neutrons from proton beams
can be a good probe of absorptive and migration effects.  We quantify how these effects modify
the Reggeized pion-exchange description of the measurements of leading neutrons at HERA.
We are able to obtain a satisfactory description of all the features of these data.
We also briefly discuss the corresponding data for leading baryons produced in hadron-hadron collisions.
\end{abstract}

\section{Introduction}

Leading neutrons, which are produced from proton beams, are of special interest since the
production process is dominated by $\pi$ exchange, see Section 2.  Such leading neutron spectra have been
measured recently in photon-proton collisions at HERA \cite{zeus1}-\cite{h12}.  These data supplement
measurements of leading baryon spectra made in hadron-hadron collisions many years ago \cite{ISRa}-\cite{ye}.
  By observing leading neutrons we have essentially
a tagged $\pi$ beam.   For example at HERA this $\pi$ beam may participate in deep inelastic scattering (DIS)
or photoproduction.  In this way both
the structure function of the pion $F_2^\pi$ \cite{h1,zeus2} and the form of the underlying
Reggeized $\pi$ trajectory \cite{zeus3} can be studied.

There is another aspect of such processes which merits particular study.
Soft rescattering effectively leads to the {\it absorption}
of leading neutrons with Feynman $x$ (which we denote $x_L$) close to 1; the role of this correction
was originally studied in Refs.~\cite{nnn,ap}.  At that time, these were essentially theoretical
studies, which provided predictions for future measurements.  Now detailed leading neutron data
have become available from the experiments at HERA.  These experiments measure the $\gamma^* p \to Xn$ cross sections
for producing leading neutrons with different $x_L$ and $p_T$ values, as functions of both the virtuality,
$Q^2$, of the photon, and the c.m. energy, $W$, of the incoming $\gamma p$ system.
The HERA data offer the opportunity of seeing how absorption changes as a function of all these
kinematic observables.  We also show the importance of {\it migration}, or change
of the kinematic variables, of the leading neutron due to rescattering effects.
The ways to estimate the effects of absorption and of migration are discussed in
Sections 3 and 4 respectively. The description of the HERA data, $\gamma p \to Xn$, is
presented in Section 5.
For completeness, in Section 6, we discuss the behaviour of leading baryons produced in
hadron-hadron collisions, in particular in the processes $pp \to Xn$ and $pn \to Xp$. 

In {\it exclusive} reactions, the rescattering could produce new secondaries which has the
effect of suppressing the rate.  That is, an exclusive cross section is
suppressed by absorptive effects.  On the other hand, for an {\it inclusive} process the rescattering,
will just change the energy and transverse momentum of the leading baryon, and will depopulate
the region of large $x_L$ and small $p_T$.  From the viewpoint of an experimental trigger, for $x_L \to 1$
this appears as absorption, but it is better to consider it as a {\it migration} into an
enlarged phase space.  In particular, such a migration can affect the form of the effective
(that is measured) $\pi$ trajectory, leading to a larger mean $p_T$ at lower $x_L$.  Of
course, this may also modify the measurement of the pion structure function, $F_2^\pi$,
which is extracted from the HERA data.

\section{Leading neutrons from $\pi$ exchange}

\begin{figure}
\begin{center}
\includegraphics[height=5cm]{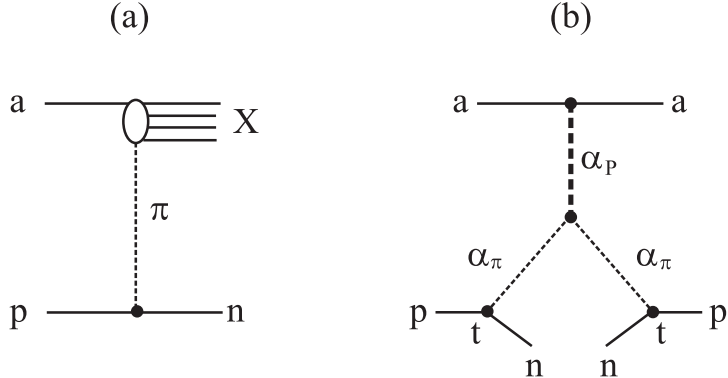}
\caption{(a) The pion-exchange amplitude and (b) the corresponding dominant triple-Regge contribution to the
cross section of the inclusive production of leading
neutrons, $ap \to Xn$.
 The coupling of the Pomeron$-\pi-\pi$ Regge trajectories is denoted by $r_{P \pi \pi}(t)$
in \eqref{eq:rrr}.\label{fig:1}}
\end{center}
\end{figure}

To begin, let us recall the triple-Regge formula which describes the inclusive processes $ap \to Xn$
with $a=\gamma$ or $p$.  For $x_L$ close to unity the inclusive leading neutron spectra is
given by the triple-Regge diagrams shown in Fig.~\ref{fig:1}. Thus we have
\be
f~\equiv~E\frac{d^3\sigma}{d^3p}~=~ {\left( \frac{s}{s_0}\right)}^{\alpha_P(0)-1}
~\frac{g^2_\pi(t)~g_P(0)~r_{P \pi \pi}(t)}{{\rm sin}^2(\pi\alpha_{\pi}(t))}~(1-x_L)^{\alpha_P(0)-2\alpha_\pi (t)},
\label{eq:rrr}
\ee
where $t=(p_p-p_n)^2$; and $g_P(0)$ is the coupling of the Pomeron to the photon, via a
$\qq$ pair, or the proton, depending on whether $a=\gamma$ or $p$.
The above form, with only the dominant pion-exchange contribution, has been used by many authors, see, for example, Refs.~\cite{zeus2,b1,b2}.
Moreover, both the old hadron-hadron interaction data and the recent HERA results on the spectra of neutrons \cite{zeus3} clearly indicate the reggeization of
pion exchange.  In fact, detailed studies of hadronic interactions in the reggeized pion-exchange
model \cite{b1,b2} have shown that the formula can be used in a broad range of $x_L$,
with $x_L \gapproxeq 0.4-0.5$.

The contribution of reggeized pion exchange to the inclusive production of baryons, $ap \to XN$ with
$N=p,n,...$, is of the form
\be
\frac{d\sigma(ap \to XN)}{dx_L dt}~=~\frac{G_{\pi^+ pn}^2~(-t)}{16\pi^2~(t-m^2_{\pi})^2}~F^2(t)~\sigma^{\rm tot}_{a\pi}(M^2)~(1-x_L)^{1-2\alpha_\pi (t)},
\label{eq:pi}
\ee
where $\alpha_\pi(t)=\alpha^\prime_\pi (t-m^2_{\pi})$ is the pion trajectory with slope $\alpha^\prime_\pi \simeq
1~\GeV ^{-2}$, and $G^2_{\pi^0 pp}/4\pi =G^2_{\pi^+ pn}/8\pi
=13.75$ \cite{GpiN}.  The invariant mass $M$ of the produced system $X$
is given by $M^2 \simeq s(1-x_L)$. The signature factor, and the difference between sin$(\pi\alpha_{\pi}(t))$ and
$(t-m^2_{\pi})$ factors in the denominators of \eqref{eq:rrr} and \eqref{eq:pi}, are absorbed in the effective
vertex form factor $F(t)$. This form factor is
usually taken to be of the form
\be F(t)~=~{\rm
exp}(bt/2),
\label{eq:F}
\ee
where, from data at relatively low energies, we expect $b \sim 4 ~{\rm
GeV}^{-2}$, see, for example, Ref.~\cite{b2}.  However more recent analyses find $b \sim 0$ \cite{kop,zeus2}. 

\begin{figure}
\begin{center}
\includegraphics[height=5cm]{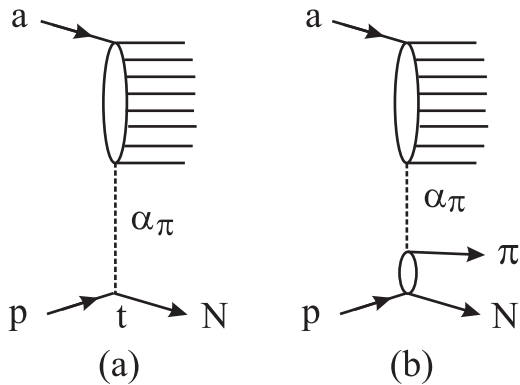}
\caption{Diagrams (a) and (b) show the reggeized $\pi$-exchange
contributions to the leading $ap \to XN$ and $ap \to X(N\pi)$ spectra
respectively, with $N=p,n$. \label{fig:2}} \end{center} \end{figure}

For the pion-exchange contributions to the spectra of both protons and neutrons, it is possible to take account,
not only of the diagram of Fig.~\ref{fig:2}a, but also, of the diagram of Fig.~\ref{fig:2}b. Besides
this we can include the contributions of $\rho,~a_2,...$ exchanges, and also of
resonance decays, such the $\Delta$-resonance contribution in the $\pi N$ amplitude, to the leading $p$ and $n$ spectra. The
calculations of the spectra of neutrons of Refs.~\cite{b1,b2,kop}, show that the diagram of Fig.~\ref{fig:2}a
dominates, and that the contribution of the diagram of Fig.~\ref{fig:2}b is about 20$\%$ at most.
On the contrary, the leading proton spectrum is dominated by the triple-Pomeron diagram; the diagram of
Fig.~\ref{fig:2}a gives only about $30\%$ of the inclusive cross section for $x_p \simeq 0.8$ in the ISR energy
range.

One interesting application of the leading neutron spectra at HERA is the possibility
 of using \eqref{eq:pi}
to extract information on the $\gamma\pi$ total cross section, $\sigma^{\rm tot}_{\gamma\pi}$, for both real
and virtual photons. This, in turn, allows a measurement of the pion structure function $F_2^\pi (x_B,Q^2)$
at very high energy or small Bjorken $x_B$. The direct application of \eqref{eq:pi} to the HERA
photoproduction data \cite{zeus2}
leads to the result  $\sigma^{\rm tot}_{\gamma\pi}/\sigma^{\rm tot}_{\gamma p}=0.32 \pm 0.03$, which is
about a factor 2 lower than the expected\footnote{Note that experimental estimates of
$\sigma^{\rm tot}_{\pi p}$ at high energies, based on the absorptive corrections to the
amplitude for $\gamma + p \to \pi^+ \pi^- +p$
and its interference with the $\gamma + p \to \rho^0 + p$ amplitude,
are about $\frac{2}{3}\sigma^{\rm tot}_{pp}$ at the same energy \cite{zeus4}. Thus we expect the
additive quark model estimates to be reliable. } ratio of 2/3.    However, so far, we have neglected
the absorptive corrections, or so-called rescattering effects,
to \eqref{eq:pi}. These are very important at high energies.  They reduce the predicted cross section, as well
as modifying both the energy and the $Q^2$
dependence of the inclusive spectra of leading neutrons.  We discuss the form of the correction in the
next Section, and present the leading neutron spectra at HERA in Section 5.

\section{Unitarity effects and gap survival}

The Born-type regge pole exchange diagram of Fig.~\ref{fig:1} represents only the first
approximation to the description of the leading neutron spectra at high energies. Multi-pomeron exchanges
introduce modifications of the amplitudes, and lead to the restoration of unitarity \cite{s2,abk}.

\begin{figure}
\begin{center}
\includegraphics[height=5cm]{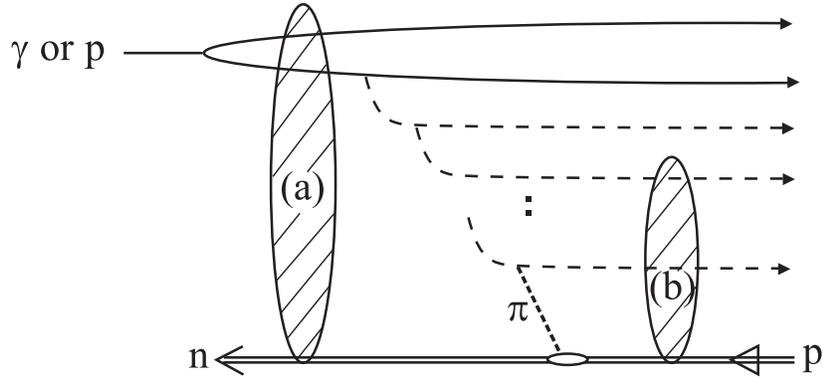}
\caption{The space-time diagram for the amplitude describing leading neutrons produced by the
inclusive processes $\gamma p \to Xn$ or $pp \to Xn$.
The two types of multi-Pomeron absorptive corrections are indicated symbolically by the
shaded areas (a) and (b).  The corresponding corrections to the cross section are shown in Fig.~\ref{fig:4}
and Fig.~\ref{fig:5} respectively.  \label{fig:3}}
\end{center}
\end{figure}
 We use the space-time diagram of Fig.~\ref{fig:3} to distinguish between two types of rescattering.
First we have the eikonal
scattering of the leading hadrons (either incoming or outgoing) indicated by exchange (a).  The other possibility is
the rescattering of the leading hadron on one of the intermediate partons in the central
rapidity region, indicated by exchange (b).
\begin{figure}
\begin{center}
\includegraphics[height=5cm]{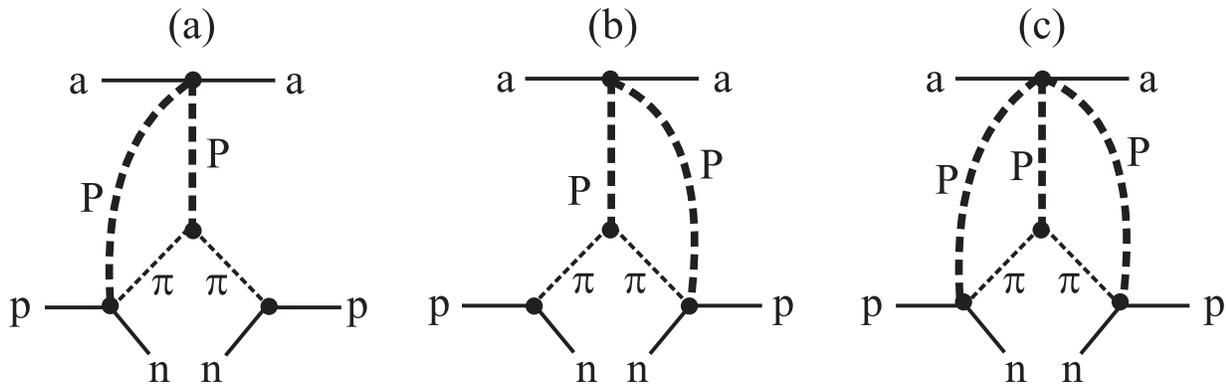}
\caption{Symbolic diagrams of the eikonal absorptive corrections to the cross
section for the inclusive process $ap \to Xn$.  The extra lines denoted by P,
which surround the triple-Regge interaction, represent multi-Pomeron exchanges
between the leading hadrons.
\label{fig:4}}
\end{center}
\end{figure}
\begin{figure}
\begin{center}
\includegraphics[height=5cm]{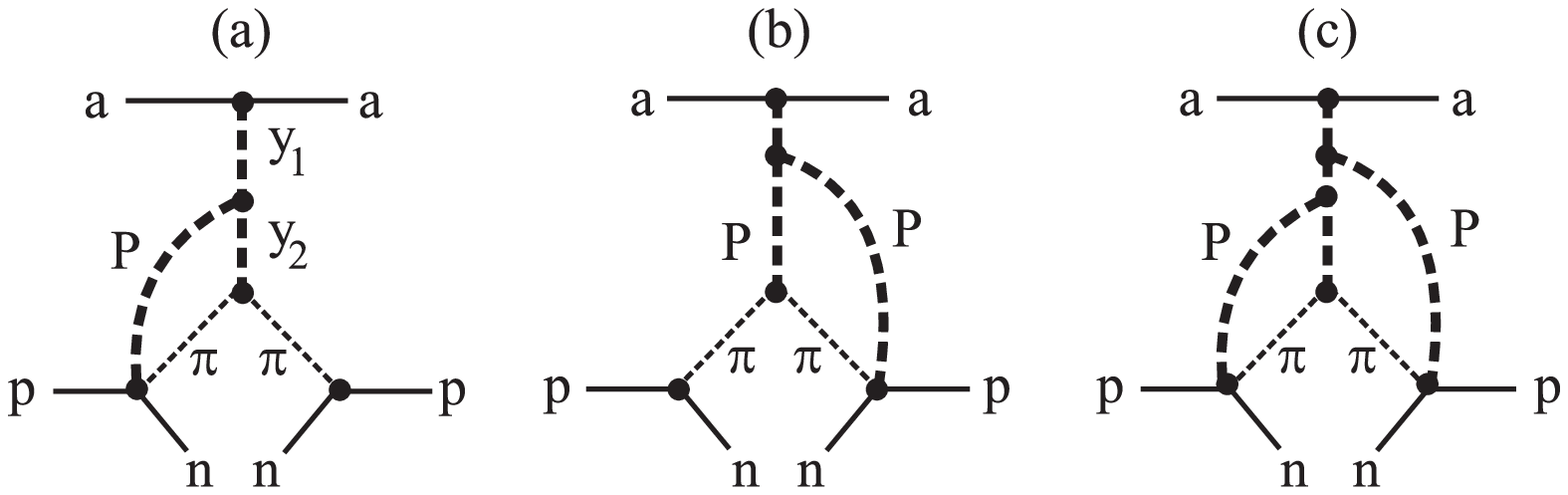}
\caption{Symbolic diagrams for the ``enhanced" absorptive corrections to the cross
section for the inclusive process $ap \to Xn$, which become 
important at very high energies.  The mass $M$ of the produced system $X$
is assumed to be sufficiently large for both the rapidity intervals $y_1$ and $y_2$ to
accommodate Pomeron exchange.  The extra lines denoted by P,
which are coupled directly to the ingoing $p$ or outgoing $n$,  represent multi-Pomeron exchanges.
\label{fig:5}}
\end{center}
\end{figure}

In terms of Feynman diagrams, an additional Pomeron exchange leads to a negative contribution;
that is absorptive corrections diminish the size of the predicted cross section.  From the physical
point of view, the inelastic interaction, shown by either exchange (a) or exchange (b) in Fig.~\ref{fig:3},
produces new secondary particles which populate the rapidity gaps and carry away energy from the
leading neutron. In other words a leading neutron, with a large energy fraction $x_L$, can only be
observed in the small fraction of events which have no secondary inelastic interactions. The rescattering corrections therefore reduce the probability that the leading
neutrons will be found in the large $x_L$ bins.
 
Technically the absorptive corrections are calculated in the following way.  The most
familiar are the eikonal-type corrections which are shown
symbolically in  Fig.~\ref{fig:4}, where only elastic
rescatterings are taken into account. The corrections are well known:
\be
d\sigma~=~\int{\rm exp}(-\Omega (s,\rho_T))~d\sigma_0(s,\rho_T...),
\label{eq:eik}
\ee
where $\rho_T$ is the impact parameter, and $d\sigma_0$ is the contribution to the cross section for
leading neutron production from the lowest order diagram of  Fig.~\ref{fig:1}(b).
The opacity $\Omega$ is the impact representation of the single-pomeron contribution to
elastic scattering, which can be determined from studies of the total and elastic processes;
for a recent review see, for example, Ref.~\cite{gotsman}.
The formalism can easily be extended to take into account diffractive excitations of the initial and
final hadrons.  It leads to a multi-channel version of \eqref{eq:eik}, see, for example
\cite{TM,s2,abk}. Diagrams of this eikonal type have been considered in previous estimates of absorptive
corrections to the spectra of leading baryons \cite{nnn,ap}.

Note that since $\Omega (s,\rho_T)$ increases with energy, the suppression of the Born cross section $d\sigma_0$
also increases with energy. The damping factor, exp$(-\Omega (s,\rho_T))$, is strongest for small values of the
impact parameter $\rho_T$. Thus the absorptive suppression is greater for the $\rho, a_2$ reggeons
than for $\pi$-exchange, which, due to the $1/(t-m^2_{\pi})$ behaviour at small $t$, has a broader distribution
in $\rho_T$.

The unitarity corrections have an interesting effect on the $t$-dependence of the cross section driven by 
pion-exchange, \eqref{eq:pi}.  As the pion pole is approached, that is as $t \ra m^2_{\pi}$, the corresponding impact parameter
becomes very large, $\rho_T \ra \infty$, and the gap survival damping factor 
\be
S^2(\rho_T) \equiv {\rm exp}(-\Omega) \ra 1.
\ee
  Thus we anticipate a more complicated $t$ dependence than that given by eqs. \eqref{eq:pi} and
\eqref{eq:F}.  First, at very small $t, ~|t| \lapproxeq 4m^2_{\pi}$, the survival factor $S^2$
rapidly decreases from $S^2=1$ at $t = m^2_{\pi}$, finally reaching a value $\sim{\rm exp}(-\Omega(\rho_t=0))$ at large $|t|$.   However before then,
the additional $t$-dependence coming from the pion vertex and trajectory reveals itself, namely $f_V(t)=$ 
exp$(bt/2)$ with $b \sim 1 ~{\rm GeV}^2$.  So, in \eqref{eq:pi}, we have
\be
F(t)=f_V(t)S(t).
\label{eq:fS}
\ee
Thus fitting the data using the single exponential form \eqref{eq:F} may confuse the situation.  Note that in the
discussion below we will use the average value of $S^2$
\be
{\hat S}^2~=~\langle{\rm exp}(-\Omega(\rho_T))\rangle.
\ee

\begin{figure}
\begin{center}
\includegraphics[height=5cm]{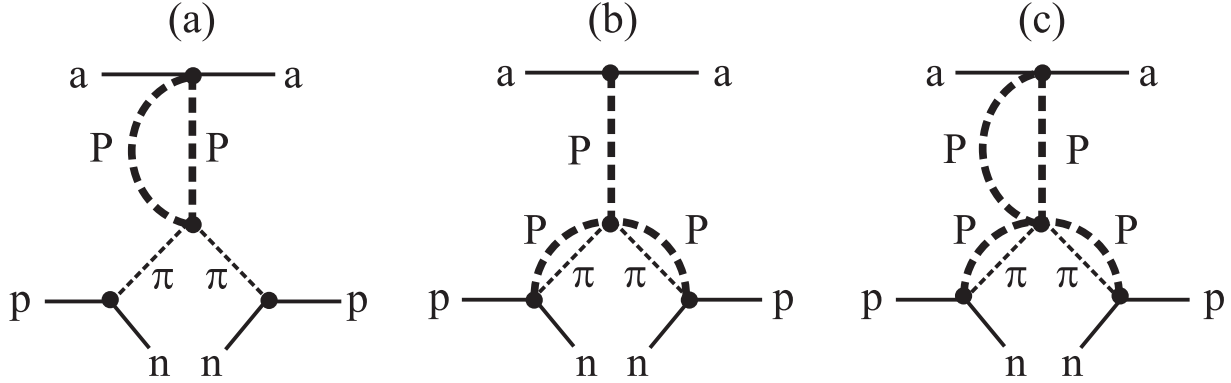}
\caption{Multi-pomeron corrections to the reggeons in the triple-regge diagrams. They have
no, or at most a weak, dependence on energy. The curved lines denoted by P represent
multi-Pomeron exchanges. \label{fig:44}}
\end{center}
\end{figure}

Another class of multi-pomeron diagrams corresponds to the modification of the upper reggeon in the triple-pomeron
diagram (Fig.~\ref{fig:44}a), or of the lower reggeons (Fig.~\ref{fig:44}b), or of both (Fig.~\ref{fig:44}c).
The diagrams of Fig.~\ref{fig:44}a modify only the $M^2$ dependence of the cross sections, and for
the $\pi$-exchange model can be absorbed in the behaviour of $\sigma^{\rm tot}_{\pi a} (M^2)$.
The diagrams of Fig.~\ref{fig:44}b modify the $x_L$ dependence, and for the $\pi$ case can be 
 by
modifying the $\pi$-trajectory (essentially the slope $\alpha^\prime_\pi)$ and the form factor.  It is important,
that to a good approximation, they do not depend on the energy\footnote{The unfactorizable
diagram  Fig.~\ref{fig:44}c can, in principle, have some weak $s$ dependence.} $\sqrt s$.

The multi-pomeron diagrams of Fig.~\ref{fig:5} are only expected to become relevant
 at sufficiently high energies. That is when the energy
is large enough for the mass $M$ of the produced system $X$ to itself be large enough for both the
rapidity intervals $y_1$ and $y_2=y_M-y_1$ to be sufficiently large to accommodate pomeron
exchange\footnote{Diagrams of this type were mentioned in our paper on dijet production at the
Tevatron \cite{s2}, but for this hard diffractive process they were small even at Tevatron energies.};
that is $y_i \ge y_0$ with $y_0 \simeq 2-3$. Recall that $y_M={\rm ln}(M^2/M^2_0)$ with $M_0^2=1~ \GeV^2$;
so for a typical value of $M^2 \sim 3000~{\rm GeV}^2$ we have $y_M=y_1+y_2 \sim 8$.
Note that rapidity intervals $y_i < y_0$ are included in the contributions of the diagrams of
Fig.~\ref{fig:4} and Fig.~\ref{fig:44}(b).

\begin{figure}
\begin{center}
\includegraphics[height=5cm]{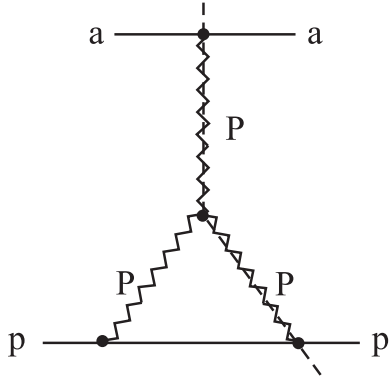}
\caption{A cut leading to an $s$-channel discontinuity of the triple-pomeron diagram for the elastic
$ap$ scattering amplitude, \label{fig:66}}
\end{center}
\end{figure}

Let us study in more detail the diagram shown in Fig.~\ref{fig:5}a, which describes the process
$ap \to Xn$. This diagram is one of the $s$-channel discontinuities
of the triple-pomeron diagram for the elastic $ap$ scattering amplitude\footnote{The summation of more complicated enhanced diagrams was discussed in \cite{schwimmer}, and more recently in \cite{BKKMR}.}, see  Fig.~\ref{fig:66}. Indeed
the diagram is related to the triple-pomeron diagram for the elastic amplitude by the AGK cutting
rules \cite{agk}. The corresponding coefficient is 4.  So it is straightforward to obtain the correction
to the $ap \to Xn$ cross section from diagram of Fig.~\ref{fig:5}a, provided we know the cross 
section of Fig.~\ref{fig:66} in the triple-Pomeron region, $x_L \to 1$.  By fitting this cross section
to the triple-Pomeron formula we can extract the strength of the vertex $r_{\rm PPP}$ \cite{trv}.  If we take the value
of the {\it effective} vertex $r_{\rm PPP}$ determined in the analysis of the old hadron data, then we find the
correction due to the enhanced absorptive diagrams to be about 15$\%$.

So there is a relatively small contribution coming from the enhanced graphs of Fig.~\ref{fig:5}, that is
from the rescattering of intermediate partons, as indicated by the shaded region (b) of Fig.~\ref{fig:3}.  This is
consistent with the fact that within the HERA range ($W = 40-270$ GeV) no energy
dependence is observed in the leading neutron yields; see, for example, Figs. 6 of \cite{h1}\footnote{See also Figs. 2 and 4 of \cite{h12} for leading proton data, where the effect of enhanced diagrams should be the same.},
and Tables 14, 18 and Figs. 11, 12 of \cite{zeus2} which show, for fixed $Q^2$, the same probability\footnote{That is
the same probability, ${\hat S}^2$, to observe the rapidity gap associated with pion-exchange.} to
observe a leading neutron for
values of $x_{Bj}$ which decrease by more than an order of magnitude corresponding to an increase of the
photon laboratory energy by more than a factor of 10.
This flat experimental behaviour of the probability indicates that actually the
absorptive corrections caused by the rescattering of intermediate
partons are much smaller than might be expected from
leading-order perturbative QCD calculations, similar to those presented
in Ref.~\cite{BM} (see also the discussion in \cite{MR,KMRext}).

\section{The inclusion of migration}
In order to compute the spectra of leading neutrons, we must consider the
effects of {\it migration}, as well as of absorption.  We assume that each
rescattering spreads out the $p_T, x_L$ spectra of the leading baryon
according to distribution\footnote{Strictly speaking, besides the
migration in $p_T, x_L$ space, there may be charge-exchange (that is neutron
to proton transitions). We neglect this effect since the goal of the
present section is not a precise quantitative description, but rather a
qualitative evaluation of the role and possible size of the `migration'
phenomena. For the same reason we use a simplified form,
 (\ref{m1}), of the leading baryon distribution.}
  \be
  \label{m1}
  \frac{dN}{dp^2_Tdx_L}\ =\ (1-a)x_L^{-a}~b_m ~{\rm exp}(-p^2_Tb_m),
  \ee
where we take the slope $b_m=6$ GeV$^{-2}$, and the intercept 
$a \sim 0 - \frac{1}{2}$ corresponding to
secondary Reggeon exchange in the Kancheli-Mueller approach.  Thus, after two rescatterings
$$
  \frac{dN^{(2)}}{dp^2_Tdx_L}\ =\ b^2_m~\int
\frac{d^2q_T}{\pi}~{\rm exp}(-q^2_Tb_m)~{\rm exp}(-(\vec p_T-\vec q_T)^2b_m)\int_0^1
\frac{dx_1dx_2}{(x_1x_2)^a}~(1-a)^2 ~\delta(x_L-x_1x_2)\ =
$$
  \be
  \label{m1a}
=\ \frac{b_m~(1-a)^2~\ln(1/x_L)}{2(x_L)^a} ~{\rm exp}(-p^2_Tb_m/2)\ .
 \ee
Correspondingly, after $k$ rescatterings the
distribution becomes
  \be
  \label{m2}
  \frac{dN^{(k)}}{dp^2_Tdx_L}\ =\  \frac{b_m~(1-a)^k~\ln^{k-1}(1/x_L)}{k(k-1)!~ (x_L)^a} ~{\rm
exp}(-p^2_Tb_m/k)~ .
\ee
The probability of
rescattering, or, equivalently, the mean number $\nu$ of rescatterings,
at fixed impact parameter $\rho_T$ is given by the opacity $\nu \equiv \Omega$
\be
\label{m3}
\Omega(\rho_T)\ =\ \frac{\sigma(s)}{4\pi B_r} ~{\rm exp}(-\rho^2_T/4B_r)\ .
\ee
Here we focus attention on migration effects in $\gamma p \to Xn$, so  
$\rho_T \equiv \rho_{\gamma N}$. The cross section $\sigma$ is, however, not
the $\gamma N$ total cross section, since the process proceeds in two stages.
First the photon fluctuates into a $\qq$ pair, which may be considered as a
sum of vector mesons, and then the vector mesons rescatter on
a nucleon.  We therefore assume $\sigma$ is the $\pi p$ total cross section
having in mind the additive quark model and/or $\rho$ meson dominance.   $B_r$ is
 the slope of the rescattering {\it amplitude} (which we take to be
$B_r=5$ GeV$^{-2}$). 

Note that the impact parameter $\rho_T$ in (\ref{m3}), which controls the
absorption and migration effects, is {\it not} equal
to the parameter $\rho_{\pi N}$ which describes the space structure of
`pure' one-pion exchange in leading neutron production. The subscript $\pi$ denotes
the `transverse' position of the $\pi-\pi-$Pomeron vertex in Fig.~\ref{fig:1}(b). That is $\rho_{\pi N}$
is Fourier conjugate of the neutron transverse momentum in the
bare (Born) amplitude leading to the cross section in (\ref{eq:pi}).
However, due to the rather large
values of $\rho_{\pi N}$, we may expect that the absorptive and
migration effects will not be too strong --- since the transverse distance between
the leading baryon and the incoming photon (or $q\bar q$-pair),
\be
\rho_T \equiv \rho_{\gamma N}~=~|\vec{\rho}_{\pi N}+\vec{\rho}_{\gamma\pi}|,
\ee
is, in turn, relatively large and the mean number of rescatterings
$\nu(\rho_T)=\Omega(\rho_T)$ at the periphery of the interaction
(described by the amplitude $\Omega$) is rather small.   Here $\rho_{\gamma\pi}$ is the
impact parameter for the amplitude describing the interaction of the incoming photon
with the `effective' pion (exchanged in the $t$-channel); recall that the subscript $\pi$ denotes
the position of the $\pi-\pi-$Pomeron vertex in Fig.~\ref{fig:4}.

\begin{figure}
\begin{center}
\includegraphics[height=15cm]{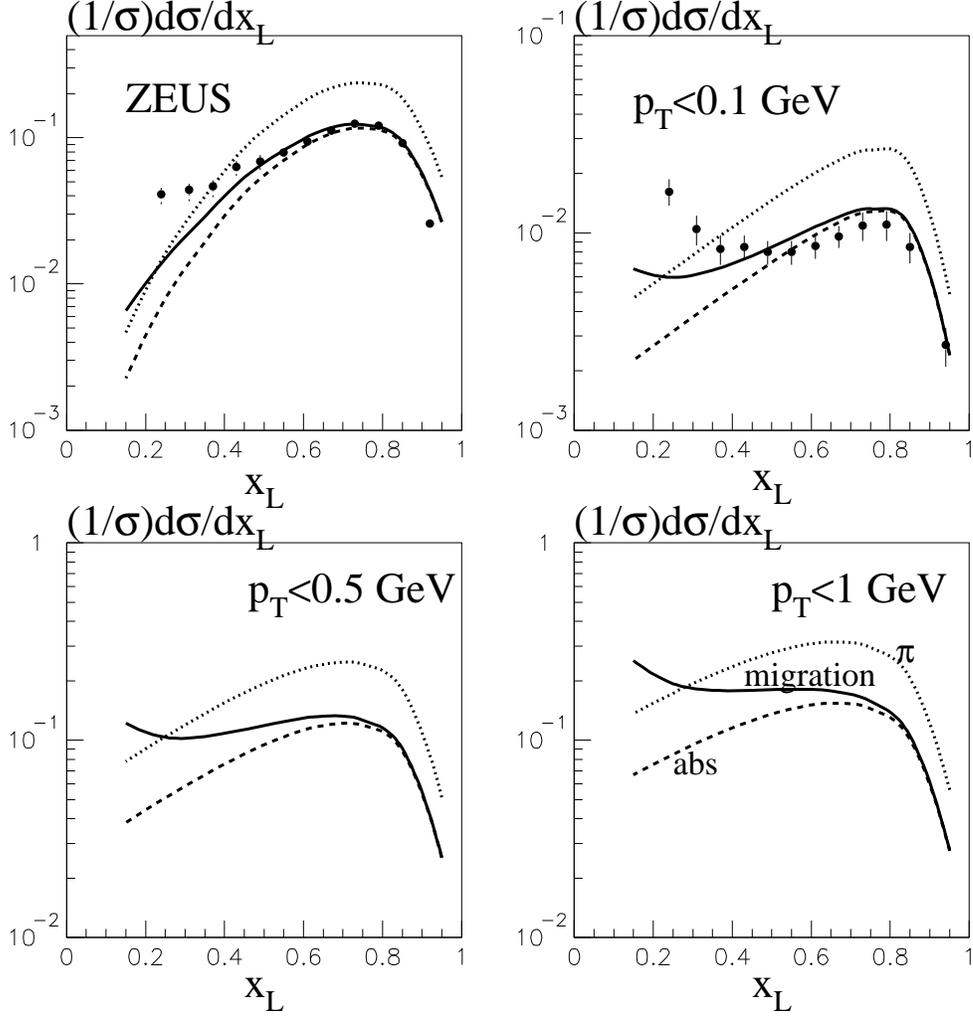}
\caption{The predictions for the $x_L$ spectra of leading neutrons
corresponding to the ZEUS kinematics.  The dotted, dashed and continuous
curves are respectively the results assuming first only Reggeised $\pi$ exchange, then including
absorptive effects, and finally allowing for migration.   The different plots show the effects of
imposing different $p_T$ cuts on the leading neutron.  The data points are the spectra obtained
from the measurements of Ref.~\cite{zeus2}. Note that the curves plotted here account only for the
pion-exchange contribution, which is not the dominant source of leading neutrons for $x_L \lapproxeq 0.4$.
\label{fig:r1}}
\end{center}
\end{figure}

To account for the fact that $\rho_T$ and $\rho_{\pi N}$ are not the same, for each value of $\rho_{\pi N}$, we
calculate the probability
\be
  \label{m5}
w(\rho_T,\rho_{\pi N})\ =\ N(\rho_{\pi N})\int
  d^2\rho_{\gamma\pi}~
\Omega_\pi(\rho_{\gamma\pi})~
  \delta(\rho_T-|\vec{\rho}_{\gamma\pi}+\vec{\rho}_{\pi N}|)\ ,
  \ee
where $\Omega_\pi(\rho_{\gamma\pi})$ is the amplitude in impact parameter space for the
photon-pion interaction. It is of the form (\ref{m3}) with
the same slope $B_{{\rm eff}}=B_r=5$ GeV$^{-2}$. Since the expression
(\ref{eq:pi}) already includes the probability of the
`effective'-pion-$\gamma$ interaction the probabilities
 $w(\rho_T,\rho_{\pi N})$ are normalized to one; that is the normalization $N(\rho_{\pi N})$ is
fixed by the condition\footnote{Strictly speaking we should work with
amplitudes and not with cross sections. Moreover the impact
parameter $\rho_T$ in the amplitude $A$ may not be equal to the
parameter $\rho^*_T$ in the conjugated amplitude $A^*$. We find
$\rho_T=\rho_T^*$ only after integration over the neutron
transverse momentun $p_T$. However to simplify the computations, here we
use the semiclassical approximation, which assumes $\rho_T=\rho^*_T$,
using the Fourier transform just for the cross section. This approach,
which is usually used in Monte Carlo simulations, may be justified by the
fact that the effect of migration becomes important only at
relatively large $p_T$, see Fig. \ref{fig:r1}.  So after we integrate up to
$p_T=0.5 - 1$ GeV we find that the equality $\rho_T=\rho_T^*$ holds to
rather good accuracy. Nevertheless, the fact that we need to correct the
normalization in (\ref{m5},\ref{m5a}) is just the result of this
semiclassical approximation.}

\be
\label{m5a}
\int d\rho_T ~w(\rho_T,\rho_{\pi N})=1
\ee
for every value of $\rho_{\pi N}$.

Thus, to calculate the inclusive cross section for leading neutron
production we have to integrate over the impact parameter $\rho_T$
\be
\frac{d\sigma}{dx_Ldp^2_T}\ =\ \int d^2\rho_T ~\frac{{\rm exp}(i\vec
p_T\cdot \vec\rho_{\pi N})} {2\pi}~d^2\rho_{\pi N}~ w(\rho_T,\rho_{\pi N})
{\cal F}(\rho_T,\rho_{\pi N},x_L)~, \label{mf1} \ee
where
\be \label{mf2}
{\cal F}(\rho_T,\rho_{\pi N},x_L)\ =\ \nu(\rho_T) e^{-\nu(\rho_T)} \left[{\cal
F}^{(0)}(\rho_{\pi N},x_L) + \int^1_{x_L} dx_L' ~{\cal F}^{(0)}(\rho_{\pi
N},x_L')~{\cal F}^{(r)}(\rho_{\pi N},x_L/x_L')\right]\ .
\ee
The factor $\nu(\rho_T) {\rm exp}(-\nu(\rho_T))$  accounts for the probability of
the first interaction and the absorption at a given impact parameter
$\rho_T$, where $\nu(\rho_T)=\Omega(\rho_T)$ is the mean number of interactions.
The first term in the square brackets, ${\cal F}^{(0)}$, is the original cross section
(\ref{eq:pi}) in the $\rho_{\pi N}$ representation, that is
$${\cal F}^{(0)}(\rho_{\pi N},x_L)\ =\   \int
\frac{d^2p_T}{2\pi}~
{\rm exp}(i\vec p_T\cdot \vec\rho_{\pi N})~\frac{d\sigma^{(0)}}{dx_Ldp^2_T}~.$$
The second term
function ${\cal F}^{(r)}$ accounts for the migration.
When we sum over $k=1,2,...$ rescatterings, the spectra
 in $\rho_{\pi n}$ representation takes the form
 $$ \frac{dN}{dp^2_Tdx_L}\ =\ \sum_{k=1}^\infty \frac{dN^{(k)}}{dp^2_Tdx_L}
\ =\ \int \frac{d^2\rho}{2\pi}~
{\rm exp}(i\vec p_T \cdot \vec\rho)~\sum_{k=1}^\infty
(\nu e^{-\rho^2/4b_m})^k~\frac{(1-a)^k~\ln^{k-1}(1/x_L)}{(k-1)!~(x_L)^a}$$
   \be \label{m4}
\ =\ \int \frac{d^2\rho}{2\pi}~{\rm exp}(i\vec p_T \cdot \vec\rho)~
\nu (1-a)  e^{-\rho^2/4b_m}~\exp\left(\nu(1-a)~e^{-\rho^2/4b_m}\ln(1/x_L)\right)~/(x_L)^a~.
\ee
Thus the function ${\cal F}^{(r)}$ in (\ref{mf2}) is
\be \label{mf5}
{\cal F}^{(r)}(\rho,x_L)\ =\ 
\nu (1-a) e^{-\rho^2/4b_m}~\exp\left(\nu (1-a)~e^{-\rho^2/4b_m}\ln(1/x_L)\right)~/(x_L)^a\ .
\ee
The results
shown below correspond to $a=0$, but the spectra obtained for $a=\frac{1}{2}$ are very similar.

\section{Predictions for leading neutrons at HERA}

To predict the absorptive corrections to the pion-exchange formula (\ref{eq:pi}), in which we set $b=0$, 
we use (\ref{eq:eik}), together with (\ref{m3}) in
which we take $\sigma=1.3 \times 31$ mb. The value $\sigma_{\rm tot}(\pi p)=31$ mb agrees
with the Donnachie-Landshoff parametrization \cite{DL} and with the
cross section evaluated by ZEUS via the $\pi$-proton absorption in
$\gamma p\to(\pi^+\pi^-)\ p$ process \cite{zeus4}. The factor 1.3 takes account of the diffractive
excitations of the initial and final hadrons \cite{KMRsoft}.

The resulting predictions for the $x_L$ spectra of leading neutrons produced at HERA are shown in
Fig.~\ref{fig:r1}.  In the first plot the ZEUS acceptance cut, $\theta_n<0.8$ mrad, has been imposed.
In the next three plots we show the effect of imposing different $p_T$ cuts, rather than
the $\theta_n$ cut, on the leading neutron.
Each plot shows three curves. The dotted curve is the Reggeised $\pi$-exchange prediction
calculated from (\ref{eq:pi}), whereas the continuous and dashed curves are computed
from (\ref{mf1}) and (\ref{mf2}) with and without migration (${\cal F}^{(r)}$) included.
The data points correspond to the leading neutron spectra measured
by ZEUS \cite{zeus2}.

As anticipated, the migration practically does not affect the
region of large $x_L>0.8$, where it is essentially enough to account for
absorption only. However at smaller $x_L$, the role of migration becomes quite important.
 First of all, it compensates the absorption (which occurs for $x_L$ close to 1), and
provides the conservation of baryon charge. Next, it changes the behaviour
of the $x_L$ and $p_T$ distributions of the leading neutrons; and modifies the
`effective' slope of the reggeized pion trajectory which was extracted from the
data using the simplified formula (\ref{eq:pi}).

From Fig.~\ref{fig:r1}, we see that the predicted leading neutron spectrum
is in satisfactory agreement with that measured by ZEUS.
The absorptive corrections reduce the cross section given simply by Reggeised pion
exchange by a factor of just less than 0.5 so that the predictions are in agreement with the
data at large $x_L$.  The reduction by 0.4-0.5 is not inconsistent with an earlier calculation \cite{KKMR034}
which gave a rapidity gap survival factor $\hat{S}^2$ of 0.34 for the resolved part
of the photon wave function.  

Moreover the data on the production of a leading neutron,
together with either a pair of high $E_T$ jets \cite{dijet}
or charm $({\rm D}^*)$ \cite{charm}, are consistent with a much larger value of $\hat{S}^2$,
as is to be expected when we select $x_\gamma \to 1$ events which sample a point-like
photon, which directly interacts with high $E_T$ dijets.  In particular, the probability to observe a
leading neutron increases by up to a factor of 2.5 (that is $\hat{S}^2$ increases
up to about 0.9) when the momentum fraction carried by the dijet grows from 0 to 1, see Fig. 8c 
of Ref.~\cite{dijet}.  In Ref.~\cite{h1dj} the H1 dijet data were described, within NLO QCD \cite{kk},
by pure pion-exchange without the inclusion of the survival factor, that is with ${\hat S}^2=1$.
However the $t$-dependence was not studied.  Moreover a larger $t$-slope parameter, $R=0.93~{\rm GeV}^{-1}$, 
was used in \cite{h1dj} than the value $R=0.5~{\rm GeV}^{-1}$ used in \cite{kk}. Thus the $b$ slope
in \cite{h1dj} is 3.5 times larger.  In this way the
absorptive corrections were mimicked in \cite{h1dj}, see the discussion leading to eq. \eqref{eq:fS}. 

The comparison with the ZEUS data in Fig.~\ref{fig:r1} shows evidence of migration for 
$0.4 \lesim x_L \lesim 0.6$.  The points at lower $x_L$ indicate that there may be additional
contributions\footnote{Note that in our computation we have already used $b=0$. Thus we have no possibility
to enlarge the neutron yield for $x_L \sim 0.5$ by diminishing the value that we take for
the slope $b$.} caused by the recombination of an initial valence quark with a sea quark in the neutron with 
$x_L \sim 0.3$, as well as by 
baryon charge transfer to the central region\footnote{The ratio $p/\bar{p} > 1$
measured in the central region at RHIC and HERA indicates the presence
of such an effect.}, whose description is beyond
the present analysis.

The leading neutron measurements at HERA also allow us to explore the $Q^2$ dependence of the problem.
Unlike the naive Vector Dominance Model (VDM), where the effect of
absorption should disappear as $1/Q^2$, in QCD (or in the Generalized
 VDM) these effects are expected to decrease as $1/{\rm ln}Q^2$.  It was shown in Ref. \cite{zeus2}
that the $Q^2$ behaviour observed at HERA is consistent with predictions \cite{nnn,ap} based on the
QCD dipole approach.

\section{Leading baryon spectra in hadron-hadron collisions}

Some years before the HERA data became available,
 leading nucleon spectra had been observed in hadron-hadron collisions in a variety of experiments, albeit at
lower energies.  Generally, these, and related, data are well described by the triple-Reggeon formalism.  At first sight, it
appears strange that no gap survival factor suppression, $\hat{S}^2$, was necessary --- particularly in view of our
discussion of the HERA data in the previous section.  However, this factor is effectively included in the
normalization of the triple-Reggeon vertices \cite{trv} which were determined by fitting the data.  The only problem
is the $\pi \pi$-Pomeron vertex, which must be consistent with the known result for the on-shell pion as
$t \to m_\pi^2$.  The on-shell vertex is known from an independent analysis of $\pi p$ total and differential
cross section data.  Now, it is crucial to note that
 the $\pi \pi $-Pomeron vertex depends strongly on the value of $t-m^2_\pi$,
leading to an effective factor $F(t)=1$ at $t=m_\pi^2$.  In the physical region of negative $t$ this factor is
expected to be $\hat{S}^2 \sim 0.4$ in the CERN ISR energy range \cite{MTU,Zak,KMRsoft}.  Indeed, the indications
for such absorptive corrections to $\pi$-exchange were observed in data 
for $pp \to X\Delta^{++}$ production \cite{Barish,MTU}.
Spectra of leading baryons in the Reggeized
 pion-exchange model have been described in papers \cite{b1,A,B,kop,MTU,Zak}. In
 papers \cite{b1,A,B} a phenomenological form-factor, which to some extent
 mimic absorptive effects, has been used.

The situation with inclusive nucleon-charge exchange in $pp$ (and $pd$)
collisions, that is in the spectra of neutrons in $pp$-collisions and protons
in $pn$-collisions, is not clear.
The largest set of the highest energy
data was obtained at the CERN ISR, Refs.\cite{ISRa} and \cite{ISRb}. These ISR spectra of neutrons
have a rather different form.  It was emphasized in Ref.~\cite{A} that the
data of \cite{ISRa} are strongly different\footnote{Note that in the CERN ISR paper \cite{ISRa} the
cross section plotted in Fig. 7 is about 3 times smaller than that given in the Tables and in Fig. 6.} from predictions of the pion-exchange 
model, and lead to a violation of the energy-momentum sum
rule, while the data of \cite{ISRb} are in a reasonable agreement
with theoretical predictions \cite{B}.  As mentioned above, the models of Refs.~\cite{A,B}
are based on Reggeized pion-exchange, with the gap
survival factor ${\hat S}^2$ represented by a form factor which, however, gave only a rather
small effect ($\sim 10\%$).  The equivalent process $pn \ra Xp$ was also
measured \cite{Robinson} over range of energies, which partly overlap with
ISR energy range and are consistent with the model of Ref.~\cite{A}. On the
other hand there are data on the process $pn \ra Xp$ at relatively low
energies (11.6 GeV), which are lower than predictions of the pion-exchange model
without the ${\hat S}^2$ factor.

Thus the final picture for hadron-hadron processes is not clear.  
Some experiments are in agreement with the theoretical expectations of
appreciable absorptive corrections to $\pi$-exchange, while others appear to be described by
simplified $\pi$-exchange.

\section{Summary}  

We have described all the main features of the leading neutrons observed at HERA, both in
photoproduction and as a function of $Q^2$.  For photoproduction we may refer to Fig.~\ref{fig:r1}. 
We see that Reggeized pion-exchange on its own is insufficient to describe the
production of leading neutrons. However the inclusion of {\it absorptive} corrections are
found to reduce the $\pi$-exchange prediction by a factor of just less than 0.5, and to bring the
theoretical expectation in line with the data.  Thus, after accounting for the rapidity gap
survival factor ${\hat S}^2$, such data can be used to measure the
$\gamma\pi$ cross section and the pion structure function.  Moreover the value of the gap
survival factor ${\hat S}^2$ may be monitored by comparing the spectra of
leading neutrons in dijet production at low $x_{\gamma}$ with those for $x_{\gamma} \to 1$,
where the cross section is dominated by the direct, point-like photon.  Next, the dependence
of the survival factor  ${\hat S}^2$ (which specifies the probability to observe a leading
neutron) on the photon energy may be used to experimentally probe the contribution of the
enhanced absorptive corrections of Fig.~\ref{fig:5}. The data appear to be flat in energy indicating only a small
contribution from enhanced diagrams. This is relevant to the proposed Higgs searches in exclusive diffractive
 production at the LHC.  Thus we conclude that the additional suppression of the exclusive Higgs cross section
caused by the enhanced diagrams cannot be as large as mentioned in Ref.~\cite{BM}, see \cite{KMRext} for a
detailed discussion. 

We also found that the neutron spectrum
shows some evidence of {\it migration} effects, see Fig.~\ref{fig:r1}.  Including migration slightly reduces the value extracted for the
slope of the pion Regge trajectory; the reduction is about $0.05-0.1 ~{\rm GeV}^{-2}$ depending
on the interval of $x_L$ used to measure $\alpha^\prime_\pi$.

Finally, in Section 6, we briefly reviewed the available experimental data for leading baryons produced in
hadron-hadron interactions, which were obtained about 30 years ago.  Here the experimental
situation is confusing.  Some experiments are in line with theoretical expectations, while
others are not.

\section*{Acknowledgements}

We thank Jon Butterworth, Albert De Roeck, Boris Kopeliovich and Graeme Watt for useful discussions.
ABK and MGR would like to thank the IPPP at the University of Durham for hospitality, and ADM thanks the Leverhulme Trust for an Emeritus Fellowship. This work was supported by the Royal Society,
the UK Particle Physics and Astronomy Research Council, by grants INTAS 00-00366, RFBR 01-02-17383 and
04-02-16073, and by the Federal Program of the Russian Ministry of Industry, Science and Technology
40.052.1.1.1112 and SS-1124.2003.2.


\end{document}